\documentclass[aps]{revtex4}

\usepackage{amsmath}
\usepackage{amssymb}
\usepackage{graphicx}
\usepackage{epsfig}

\newcommand{\dd}{\,{\rm d}}
\newcommand{\ii}{{\rm i}}
\newcommand{\e}{{\rm e}}

\begin{document}

\title{Influence of  modal loss on the quantum state generation via
  cross-Kerr nonlinearity}

\author{D. Mogilevtsev$^{1,2}$}\author{Tom\'a\v{s} Tyc$^{3,4}$} \author{N. Korolkova$^3$}
\affiliation{$^1$Institute of Physics, Belarus National Academy of
Sciences, F.Skarina Ave. 68, Minsk 220072 Belarus \\
$^2$Instituto de F\'{\i}sica, UNICAMP, CP 6165, Campinas - SP,
13083-970, Brazil \\
 $^3$School of Physics and Astronomy, University of St
Andrews, North Haugh, St Andrews KY16 9SS, UK \\
$^4$ Institute of Theoretical Physics, Masaryk University,
Kotlarska 2, 61137 Brno, Czech Republic}

\begin{abstract}
In this work we investigate an influence of decoherence effects on
quantum states generated as a result of the cross-Kerr nonlinear
interaction between two modes. For Markovian losses (both photon
loss and dephasing), a region of parameters when losses still do
not lead to destruction of non-classicality is identified. We
emphasize the difference in impact of losses in the process of
state generation as opposed to those occurring in propagation
channel. We show moreover, that correlated losses in modern
realizations of schemes of large cross-Kerr nonlinearity might
lead to enhancement of non-classicality.
\end{abstract}


\date{Version 6, changed by Natasha, February 2009}

\maketitle


\section{Introduction}

Nowadays cross-Kerr nonlinearity is considered as a promising tool
for quantum computation and non-classical state generation
\cite{kock2007}. An entanglement arising between modes
participating in the cross-Kerr interaction can be used for
establishing an interface between matter qubits and `flying' light photonic
\cite{lee2006}, for generation of non-Gaussian states and
Schr\"odinger-cat states \cite{{Tyc08},{clancy2008}}, and for performing
quantum gate operations \cite{spiller2005}. An interest to
cross-Kerr nonlinear interactions is heated up by both
developing ways to implement effectively even very weak
non-linearities (which one commonly expects to have in practice)
\cite{{spiller2005},{rohde2008},{Jeong06}},  and by discovery of
methods to produce sufficiently large self-Kerr and cross-Kerr
nonlinearities (such as implementation of the electromagnetically
induced transparency (EIT)
\cite{{Fleischhauer},{korolkova2007},{plenio08}} and photonic
crystals \cite{vuckovic}).

Decoherence is a main practical obstacle to implementations of
schemes using Kerr and cross-Kerr nonlinearities. A genuine
example of quantum state degradation due to losses is a
decoherence of a quantum superposition state, the effect
drastically enhanced if speaking of a macroscopic superposition
state of the Schr\"odinger cat type. Already more than 20 years
ago the self-Kerr nonlinearity was proposed as a tool for
generating such a Schr\"odinger-cat state (more precisely,  a
superposition of two coherent states with the same amplitudes but
opposite phases) \cite{yurke}. However, photon losses turn this
superposition into statistical mixture of two coherent states with
the rate proportional to the square modulus of the amplitude of
these states. Modifications of the scheme for the cross-Kerr
nonlinearities or four-wave mixing brought no advantage with
respect to photon loss \cite{clancy2008}. This unfortunate
circumstance made one look for the ways to circumvent the problem
of decoherence that inevitably accompanies Kerr nonlinearity.
Recent suggestions in this direction are based on the conditional
preparation of desired states (which brings into consideration an
additional problem of the finite detection efficiency), and are
aimed to exploit weak nonlinearities
\cite{{spiller2005},{Jeong06},{clancy2008}}. Recently, even a way
to produce cat-states `on demand' was suggested using a source of
single-photons \cite{kim2007}.

In our work we want to discuss an aspect of the decoherence which
has been seldom discussed when considering an influence of losses on
states generated via Kerr nonlinearity. Namely, we address losses
arising \textit{in the process} of generation and not due to propagation of the generated state via lossy
channels. We concentrate our attention on a feature that might
be quite significantly pronounced in modern schemes of
generating large Kerr nonlinearity:
the  modal loss can be strongly correlated. Indeed, the modes occupy the
same volume and interact with the same physical systems which form
the reservoirs. Also, if the Kerr-nonlinearity scheme implies a
sufficiently strong dispersive coupling of light modes to
emitters, then coupling of these emitters to dissipative reservoirs
might also appear to be quite strong. As a result, this would mean
strongly correlated modal losses. For example, in photonic crystals high
density of states in the vicinity of a modal frequencies and
emitter's transition frequency can cause the strong
emitter-field coupling; but it would also imply higher
population loss of emitters due to coupling to radiative
reservoirs. Dephasing losses of emitters would as well invoke a
correlated modal dephasing.

Coupling to correlated reservoirs can drastically change state
dynamics in comparison with loss to uncorrelated reservoirs. For
example, it was demonstrated that coupling to the common reservoir
preserves entanglement of a two-mode state \cite{prauzner}.
Moreover, coupling to the common reservoir is capable of creating an
entanglement between states of initially unentangled modes even in
absence of any direct interaction between them
\cite{{braun},{horhammer}}.

In our work we demonstrate both how the correlated loss arises via
Kerr nonlinear process, and how it affects the generated states. For
this purpose we derive analytic solution generalizing a powerful
and illustrative method of Chaturvedi and Srinivasan \cite{Cha91}.
On a number of examples we show how the correlated loss enhances
and creates intermodal correlations and even entanglement, and
might lead to generation of entangled states quite different from
those generated in the same scheme without loss. Correlated loss
can result in the significantly enhanced robustness of the
generation scheme.

The outline of the paper is as follows. In Sec. II  and in the related Appendices we
describe how cross-correlation terms emerge via correlations of
Markovian reservoirs; we consider an example of
the emitter-field interaction schemes producing correlated modal
losses in the Section III. Then in Sec. IV we describe the method for obtaining exact
solutions of the cross-Kerr nonlinear interaction between modes in
presence of losses to uncorrelated reservoirs and give
generalization of the method for some cases of correlated losses.
In the Sec. V we analyze influence of losses in the
nonclassical state generation process for the case of uncorrelated
loss. Some examples of correlated losses are
considered in Sec. VI.


\section{Master equation for correlated and uncorrelated loss}
\label{master_section}

To illustrate clearly an influence of correlated and uncorrelated
losses, we restrict ourselves to the Markovian loss accounting for photon losses and dephasing of
interacting modes. We start from the general effective Hamiltonian $H(t)$
describing both self- and cross-interaction (for the moment we refrain
from detailing it) and
interaction of modes with reservoirs responsible for losses $V_{\rm loss}(t)$:
\begin{eqnarray}
V(t)=H(t)+V_{\rm loss}(t), \label{ham_ger}
\\
V_{\rm loss}(t)=a_1^{\dagger}\Gamma_1(t)+\Gamma_1^{\dagger}(t)a_1+a_2^{\dagger}\Gamma_2(t)+\Gamma_2^{\dagger}(t)a_2+
a_1^{\dagger}a_1 D_1(t)+a_2^{\dagger}a_2D_2(t). \label{vloss}
\end{eqnarray}
Here we use the interaction picture with respect to the
free Hamiltonians of reservoirs and the modes participating in the
interaction process. These modes are described by usual bosonic
creation and annihilation operators satisfying
\[ [a_1,a_1^{\dagger}]=[a_2,a_2^{\dagger}]=1, \quad
[a_1,a_2^{\dagger}]=[a_1,a_2]=0.
\]
The operators $\Gamma_{1,2}(t)$ and $D_{1,2}(t)$ describe
reservoirs responsible, correspondingly, to the photon losses in
modes $a_1$ and $a_2$, and to dephasing of these modes. They may
include also stochastic variables describing different
realizations of reservoirs.

It should be emphasized that reservoir operators $\Gamma_{1,2}(t)$
and $D_{1,2}(t)$ (together with the initial state of the
reservoir) completely describe the reservoir properties with
respect to the interaction with the modes. These operators are
built on the basis of underlying microscopic model and account for
all relevant physical parameters. For example, if the photon loss
reservoir of the first mode is composed of electromagnetic field
modes with frequencies $w_j$, described by the creation and
annihilation operators $b_j, b_j^{\dagger}$, then
\begin{equation}
\Gamma_1(t)=\sum\limits_jg_jb_j\exp\{-iw_jt\}, \label{gamma1}
\end{equation}
where each $g_j$ is the constant of interaction of the mode $a_1$
with the $j$th mode of the reservoir. Sets of frequencies $w_j$
and interaction constants $g_j$ describe completely physical
properties of the reservoir. In particular, if the reservoir is
the set of electromagnetic modes of a non-absorbing dielectric
structure, they are found from the eigensolutions of Maxwell's
equations for this structure \cite{loudon}.

We make a number of standard assumptions in considering
decoherence: coupling with the reservoirs is weak, the initial
state of the modes $a_j$ are uncorrelated with initial state of
reservoirs, and correlation times of reservoirs are small enough
to enable an implementation of the Born-Markov approximation.
Then, using, for example, a time-convolutionless projection
operator technique \cite{breuer}, one can obtain the master
equation, the Liouville equation of the following form:
\begin{eqnarray}
\nonumber {\dd\over \dd t}\rho(t)=-\ii[H(t),\rho(t)]+ {1\over 2
}[(\gamma_1-\gamma_{12})\mathcal{L}(a_1)+
(\gamma_2-\gamma_{12})\mathcal{L}(a_2)+
(d_1-d_{12})\mathcal{L}(a_1^{\dagger}a_1)+(d_2-d_{12})\mathcal{L}(a_2^{\dagger}a_2)]\rho(t)+\\
{1\over 2
}[\gamma_{12}\mathcal{L}(a_1+a_2)+d_{12}\mathcal{L}(a_1^{\dagger}a_1+a_2^{\dagger}a_2)]\rho(t),
\label{master_markov}
\end{eqnarray}
where the superoperator $\mathcal{L}(b)$ acts on the density
matrix as
\begin{eqnarray}
\mathcal{L}(b)\rho(t)=2b\rho(t)b^{\dagger}-b^{\dagger}b\rho(t)-\rho(t)b^{\dagger}b.
\label{diagonal form}
\end{eqnarray}
Here we set $\hbar= 1$ for simplicity. For more details of the derivation see Appendix A.
By construction, the master equation (\ref{master_markov}) provides for non-negative definite $\rho(t)$
for arbitrary $t\geq0$.

Note that we call reservoirs "correlated", if their integrated
cross-correlation function is non-zero, for example, \[\int
d\tau\langle \Gamma_1^{\dagger}(t)\Gamma_2(\tau)\rangle_{\rm
r}\neq 0\] and, correspondingly, the coefficient $g_{12}$ defined
in Eq. (\ref{gamma12}) is non-zero. Of course, one can always
transform Eq. (\ref{master_markov}) to the diagonal form. However,
in this case Lindblad operators (i.e. operators like $b$ in the
diagonal form (\ref{diagonal form})) will be the linear
superpositions of the former Lindblad operators, and the
transformed equation will be still describing a coupling between
physical objects represented by these original Lindblad operators
(modes $a_{1,2}$ in our case). For example, the possibility that
both modes are coupled to the same reservoir (either the photon
loss reservoir or the dephasing one) corresponds to an equality in
relations (\ref{rates_unequalities}) in Appendix A. In this case
the operators describing the reservoir are proportional to each
other, say, $\Gamma_1(t)=x\Gamma_2(t)$ and $D_1(t)=yD_2(t)$.
Equation (\ref{master_markov}) then reduces to
\begin{eqnarray}
\nonumber {\dd\over \dd t}\rho(t)=-\ii[H(t),\rho(t)]+ {1\over 2
}[\gamma_{1}\mathcal{L}(a_1+xa_2)+d_{1}\mathcal{L}(a_1^{\dagger}a_1+ya_2^{\dagger}a_2)]\rho(t).
\label{master_markov_corr}
\end{eqnarray}
The most important point here is that in this equation both modes
behave like a single object with respect to relaxation. Such a
`decoherence' is able to induce enduring entanglement between
modes $a_1$ and $a_2$ (an example is shown in the Appendix B).

\begin{figure}[th]
\centerline{\psfig{file=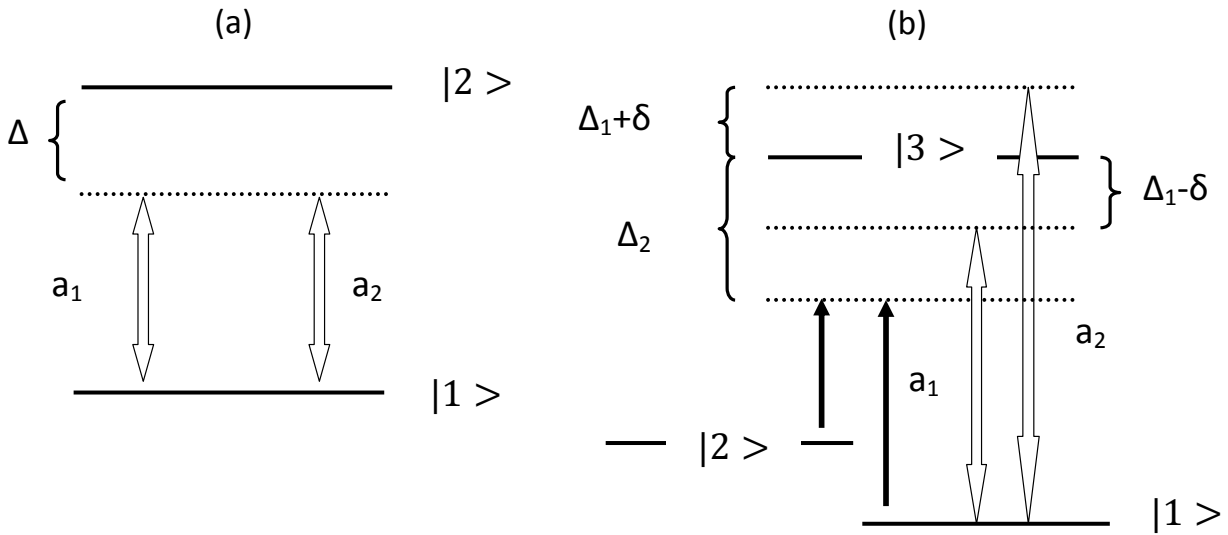,width=\linewidth}} \vspace*{8pt}
\caption{Examples of level structures for generating cross-Kerr
nonlinearities. Coupling to the quantized modes is shown by thin
arrows and coupling to classical driving fields is shown by thick
arrows. } \label{fig1}
\end{figure}

Furthermore, interacting light modes might
experience correlated losses due to fact that they both interact with the same atoms.
One can illustrate the mechanism of such a loss by the
following qualitative consideration. Let us consider a general
Hamiltonian describing the light-atom interaction $H_0$ plus
terms describing interaction of atoms with the dissipative
reservoirs
\begin{eqnarray} H_{\rm total}=H_0(t;a_1,a_2,S_k)+\sum\limits_j S_j\Gamma_j, \label{ham-loss}
\end{eqnarray}
where $S_j$ and $\Gamma_j$ are operators describing the atoms and
dissipative reservoirs, respectively. The Hamiltonian of this form describes a general interaction between field modes and emitters of some type as the correlated loss can occur in different physical systems.  Provided atom-reservoir
interactions in (\ref{ham-loss}) is sufficiently weak as not to perturb much interaction between light modes and atoms,
the following approximation can be used for the atomic operators:
\[S_j(t)\approx U^{\dagger}(t)S_j(0)U(t)\approx
F_j(t;a_1,a_2,S_k(0)),
\]
where $U(t)=\mathrm{T}\exp\left\{ -\ii\int\limits_0^t\dd\tau H_0(\tau;
a_1,a_2,S_k)\right\}$, and T denotes the time-ordering operator.
After averaging out the atomic variable, the terms that describe reservoir-mode coupling in the effective interaction Hamiltonian will take the form of
$\sum\limits_j f_j(t;a_1,a_2)\Gamma_j$. Here
$f_j(t;a_1,a_2)=\langle F_j(t;a_1,a_2,S_k(0))\rangle_{\rm am}$
and $\langle\ldots\rangle_{\rm am}$ denotes the averaging over
atomic states (in general, over emitter's states). Thus, one can see that coupling of light modes to the same atom
(emitter) interacting with the dissipative reservoir under the
condition of adiabatic elimination of emitter's variables leads
directly to the mode-reservoir interaction terms in the
resulting effective Hamiltonian. Note, that these terms remain
linear in reservoir operators $\Gamma_j$. Hence one can derive a
master equation averaging over the reservoir in a standard way.
In the Appendix B we give examples of
the derivation of effective Hamiltonians and the corresponding
master equations discussing the simplest two-mode Jaynes-Cummings
system (Fig.~\ref{fig1}(a)). The described scenario of how
correlated loss emerge is quite general, and can take place for a
wide range of schemes involving light-shift-induced photonic
nonlinearities.

A three-level
$\Lambda$-system interacting with classical driving and quantum
fields represent a more realistic example of the scheme with the correlated loss.
\label{lambda-system}
Consider the large cross-Kerr
nonlinearity generation suggested in Ref. \cite{plenio08} and
depicted in Fig.~\ref{fig1}(b). There two modes are coupled to the
transition between 1 and 3 levels of the $\Lambda$-system in
presence of two classical driving fields on transitions 1-2 and
1-3. In the rotating-wave approximation and in the interaction
picture with respect to free Hamiltonian of the reservoir, the
problem is described by the following interaction Hamiltonian:
\begin{eqnarray}
H_{\rm eff}=
[g_1a_1^{\dagger}\exp\{\ii(\Delta_1-\delta)t\}+g_2a_2^{\dagger}\exp\{\ii(\Delta_1+\delta)t\}+
\Omega\exp\{\ii\Delta_2t\}+\Gamma_{13}(t)]\sigma_{13}+\Omega\exp\{\ii\Delta_2t\}\sigma_{23}+\mbox{H.
c.}, \label{hamiltonian_lambda_system}
\end{eqnarray}
where $\Omega$ is the Rabi frequency of the driving fields;
$\sigma_{kl}=|k\rangle\langle l|$, $k,l=1,2,3$; and $g_{1,2}$ are the
interaction constants for the coupling of a light mode to an emitter (atom).  Here for simplicity we
have taken into account only losses on the transition 1-3. The
setup depicted in Fig.~\ref{fig1}(b) can be realized in toroidal
microcavities, where $a_1$ and $a_2$ correspond to the clockwise
and counter-clockwise propagating modes \cite{kippenberger}. As
usually, we assume the Markovian reservoir and the following conditions hold:
\[ \langle\Gamma_{13}(t)\rangle_{\rm r}=0, \quad
\langle\Gamma_{13}(t)\Gamma_{13}^{\dagger}(\tau)\rangle_{\rm r}=\gamma\,
\delta(t-\tau).
\]
From the Hamiltonian (\ref{hamiltonian_lambda_system}), the dynamics is governed by
\begin{eqnarray}
{\dd\over \dd t}\sigma_{31}\approx
\ii[g_1a_1^{\dagger}\exp\{\ii(\Delta_1-\delta)t\}+g_2a_2^{\dagger}\exp\{\ii(\Delta_1+\delta)t\}+
\Omega\exp\{\ii\Delta_2t\}](\sigma_{33}-\sigma_{11})+\ii\Omega\exp\{\ii\Delta_2t\}\sigma_{21}.
\label{31equation_lambda_system}
\end{eqnarray}
Using the approach described in Ref. \cite{plenio08}, we assume
that level 3 remains practically unpopulated,
$g_k/(\Delta_1\pm\delta)\ll 1$ and $\Omega/\Delta_2 \ll 1$,
as well as $g_{k}, \Omega \ll |\Delta_1-\Delta_2|$. Thus the
$\Lambda$-system is prepared in the superposition of the
metastable levels 1 and 2 (namely, in the state
$(|1\rangle-|2\rangle)/\sqrt{2}$). Then, neglecting small and
rapidly oscillating terms, one obtains from
Eqs.(\ref{hamiltonian_lambda_system}, \ref{31equation_lambda_system})
the following master equation
\begin{eqnarray}
\nonumber {\dd\over \dd t}\rho(t)\approx -{i\Delta_2\over
2\Omega^2} \left[\left({g_1^2\over\Delta_1-\delta}
a_1^{\dagger}a_1+ {g_2^2\over\Delta_1+\delta}
a_2^{\dagger}a_2\right)^2,\rho(t)\right]+{1\over2}\left[(\gamma_1-\gamma_{12})
\mathcal{L}(a_1)+(\gamma_2-\gamma_{12})\mathcal{L}(a_2)\right]\rho(t)+ \\
{\gamma_{12}\over2}\mathcal{L}(a_1+a_2)\rho(t),
\label{master_lambda_model}
\end{eqnarray}
where \[\gamma_{1,2}=\gamma{g_{1,2}^2\over (\Delta_1\mp\delta)^2},
\quad \gamma_{12}=\gamma {g_1g_2\over (\Delta_1^2-\delta^2)}.
\]

The master equation (\ref{master_lambda_model}) describes
the cross-Kerr and self-Kerr interactions of two light modes plus
their coupling to the  correlated reservoirs. As
follows from the model depicted in Fig.~\ref{fig1}(b)), it is actually the
same reservoir: one can see that $\gamma_1\gamma_2=\gamma_{12}^2$.
Naturally, loss rates ${\gamma_{1,2}}$ are much less than the loss
rate of the emitter. However, if one deals with input
modes in a coherent state of rather large amplitude for generating
large cross-Kerr nonlinearities, then even comparatively
small losses can strongly influence mode dynamics. Below we
consider examples of such an influence, e.g.,  an example of a
single-mode Sch\"{o}dinger-cat state.

The occurrence  of a correlated modal loss
due to presence of the emitter (atom in our case) has been noticed in Ref.
\cite{plenio08}. They have also pointed out that the loss rate
should be proportional to the square of the ratio of the
mode-emitter interaction constant and the detuning. However, we
would like to emphasize that this loss rate does not
depend on the population of level 3, as it can be seen from Eq.~(\ref{31equation_lambda_system}).


\section{Cross-Kerr interaction model}

Now let us turn to the specific nonlinear interaction described by the same type of the Hamiltonian as in the examples
above and in the Appendix B. We will consider the effective Hamiltonian in the master equation~(\ref{master_markov}) $H(t)\equiv H_0$ in the following general form
\begin{eqnarray}
H_0=  \sum\limits_{k,l=1}^2\chi_{kl}
a_k^{\dagger}a_ka_l^{\dagger}a_l \label{h_0_kerr}.
\end{eqnarray}
It describes the cross-Kerr and self-Kerr interaction with nonlinear coefficients $\chi_{kl}$ of two modes (or
mode superpositions) $a_1$ and $a_2$. To solve the
problem described by the master equation (\ref{master_markov})
with the Hamiltonian (\ref{h_0_kerr}), we adopt a simple and
illustrative `thermofield' notation  \cite{Ume82, Cha91}.
Essentially, instead of a density matrix acting on some space
$\mathcal{H}$, say,
$\rho=\sum\limits_{k,l}\rho_{kl}|k\rangle\langle l|$, where
$|k\rangle$ is the Fock  state with $k$ photons in $\mathcal{H}$,
we consider a state vector
$|\rho\rangle=\sum\limits_{k,l}\rho_{kl}|k\rangle {|\tilde
l\rangle}$ in an extended space
$\mathcal{H}\bigotimes\mathcal{H}^{\ast}$, where ${|\tilde
l\rangle}$ is the Fock  state with $l$ photons in
$\mathcal{H}^{\ast}$. So when the mode operators $a$ and
$a^{\dagger}$ in the master equation (\ref{master_markov}) act on
the density matrix from the left, one introduces operators
${\tilde a}$ and ${\tilde a}^{\dagger}$ is such a manner that
\[|k\rangle\langle l|a\longrightarrow {\tilde a}^{\dagger}|k\rangle {|\tilde
l\rangle}, \quad |k\rangle\langle l|a^{\dagger}\longrightarrow
{\tilde a}|k\rangle {|\tilde l\rangle}.
\]
Obviously, operators $a$ and $a^{\dagger}$ commute with ${\tilde
a}$ and ${\tilde a}^{\dagger}$. An action of the  superoperator
$\mathcal{L}(a)$ on the density matrix can be represented in the
thermofield notation as
\[
\mathcal{L}(a)\rho\longrightarrow
\mathbf{L}(a)|\rho\rangle=(2a\tilde a-a^\dagger a-\tilde
a^\dagger\tilde a)|\rho\rangle.
\]
Also, the commutator of any function of the operators $a$ and
$a^{\dagger}$, e.g., the Hamiltonian $H_0(t;a,a^{\dagger})$ is re-written as:
\[
[H_0(t;a,a^{\dagger}),\rho]\longrightarrow
\mathbf{H}_0(t)|\rho\rangle=\left(H_0(t;a,a^{\dagger})
   -H_0(t;{\tilde a},{\tilde a}^{\dagger})\right)|\rho\rangle.
\]
With help of these notations Eq.~(\ref{master_markov}) can be
represented in the `Hamiltonian' form as
\begin{eqnarray}
\nonumber {\dd\over
\dd t}|\rho(t)\rangle&=&\mathbf{H}_{\rm total}|\rho(t)\rangle\equiv \\\nonumber
&&{1\over 2 }(-2\ii\mathbf{H}_0(t)
+(\gamma_{a_1}-\gamma_{12})\mathbf{L}(a_1)+(\gamma_{2}-\gamma_{12})\mathbf{L}(a_2)+
(d_{1}-d_{12})\mathbf{L}(a_1^\dagger
a_1)+(d_{2}-d_{12})\mathbf{L}(a_2^\dagger a_2))|\rho(t)\rangle \\
&+& {1\over 2
}(\gamma_{12}\mathbf{L}(a_1+a_2)+d_{12}\mathbf{L}(a_1^\dagger
a_1+a_2^\dagger a_2))|\rho(t)\rangle, \label{master_markov_term}.
\end{eqnarray}
Its solution is then of the form
\begin{eqnarray}
|\rho(t)\rangle=\exp\{\mathbf{H}_{\rm total}t\}|\rho(0)\rangle.
\label{general solution}
\end{eqnarray}

The advantage of using thermofield notation over more traditional
algebraic manipulation with superoperators is that in many
situations (and, particularly, ones of our interest) it enables to
simplify, make more illustrative and less cumbersome finding the
solution (\ref{general solution}) and estimation of time-dependent
matrix elements. In particular, it allows to represent in a simple
form a factorization of the superoperator $\exp\{\mathbf{H}_{\rm
total}t\}$ into multipliers with easily estimated actions on the
number states \cite{Cha91}.

To illustrate this, let us consider
a simple problem of modal loss in a single mode described by the equation
\begin{eqnarray}
\nonumber {\dd\over \dd t}|\rho(t)\rangle={1\over 2
}\gamma_{a}\mathbf{L}(a)|\rho(t)\rangle. \label{master equation
example}
\end{eqnarray}
The key to solving this equation lies in the observation that the
operators
\begin{eqnarray}
A_+\equiv a^\dagger\tilde a^\dagger, \quad A_-\equiv a\tilde a,
\quad A_3\equiv (a^\dagger a+\tilde a^\dagger\tilde a+1)/2
\label{algebra definition}
\end{eqnarray}
generate the SU(1,1) algebra with the Casimir invariant $A_0\equiv
a^\dagger a-\tilde a^\dagger\tilde a$. Using the disentangling
theorem \cite{Wod85} for this group, we arrive at the simple
result
\begin{eqnarray}
  |\rho(t)\rangle =\exp\left\{
  \frac{\gamma_a}2t\right\}
  \exp[-\gamma_atA_3]
  \exp[(1-\e^{-\gamma_at})A_-]|\rho(0)\rangle.
\label{solution example}
\end{eqnarray}
Generally, the dynamics described by the solution (\ref{solution
example}) leads to transforming initially pure states into
mixtures. However, for a coherent initial state with the
amplitude $\alpha$,
\[|\rho(0)\rangle=|\alpha\rangle|\tilde \alpha^*\rangle=\sum\limits_{m,n=0}^{\infty}\exp\{-|\alpha|^2\}
\frac{\alpha^m\alpha^{*n}}{\sqrt{m!n!}}|m\rangle|\tilde n\rangle,
\]
equation~(\ref{solution example}) gives the following result:
\[|\rho(t)\rangle=|\alpha \exp\{-{\gamma_at}/2\}\rangle|\overline{
\alpha^*\exp\{-{\gamma_at}/2}\}\rangle.
\]
Returning now to the effective Hamiltonian (\ref{h_0_kerr}) that describes
cross-Kerr and self-Kerr interaction of two modes (or modal superpositions)
$a_1$ and $a_2$, we write it in the thermofield notation as follows:
\begin{eqnarray}
H_0 \longrightarrow \mathbf{H}_0=
\sum\limits_{k,l=1}^2\chi_{kl}A_0^{(k)}(2A_3^{(l)}-1).
\label{h_0_kerr_general}
\end{eqnarray}
Operators in Eq.
(\ref{h_0_kerr_general}) are $A_0^{(k)}=a_k^{\dagger}a_k-\tilde
a_k^\dagger\tilde a_k$, $A_3^{(k)}=(a_k^{\dagger}a_k+\tilde
a_k^\dagger\tilde a_k+1)/2$.

For completely uncorrelated reservoirs of different modes (i.e.
$\gamma_{12}=d_{12}=0$) the master equation
(\ref{master_markov_term}) with the Hamiltonian $H_0$ given by
Eq.~(\ref{h_0_kerr_general}) can be solved exactly using the
following factorization \cite{Cha91}:
\begin{eqnarray} \nonumber
\exp\{\mathbf{H}_{\rm total}t\}=\exp\left\{\left[-\frac{d_1}2\,(A_0^{(1)})^2-\frac{d_2}2\,(A_0^{(2)})^2
  +\ii p_1A_0^{(1)}+\ii p_2A_0^{(2)} +\frac{\gamma_1+\gamma_2}2\right]t\right\} \\
\times
  \exp[(\ii P_1-\gamma_1)A_3^{(1)}t] \exp[(\ii P_2-\gamma_2)A_3^{(2)}t]
  \exp[\Gamma_{1-}A_-^{(1)}] \exp[\Gamma_{2-}A_-^{(2)}]
\label{ham}
\end{eqnarray}
where superoperators $A_{-}^{(1,2)}$  are defined similarly as in
Eq.~(\ref{algebra definition}) and
\begin{eqnarray}
\nonumber p_k=\sum\limits_{l=1}^2\chi_{kl}, \quad \Gamma_{k-}=
\frac{\gamma_k[\e^{(\ii P_k-\gamma_k)t}-1]}{\ii P_k-\gamma_k},
\quad P_k=2\sum\limits_{l=1}^2\chi_{kl}A_0^{(l)}.
\end{eqnarray}
It is useful to note that in Eq. (\ref{ham}) all multipliers apart
from two last ones are diagonal in the number-state basis.  Also,
it is easy to see that $[\Gamma_{k-},A_-^{(k)}]=0$, operators
$\Gamma_{k-}$ are diagonal in the number-state basis, and
operators $A_-^{(k)}$ are simply products of annihilation
operators. Thus, Eq. (\ref{ham}) provides for simple analytic
solutions both for coherent initial states of interacting modes.

The solution (\ref{ham}) for uncorrelated reservoirs can be
straightforwardly generalized for some special cases of
correlated reservoirs and the Hamiltonian (\ref{h_0_kerr}). For
example, let us consider the problem without dephasing,
$d_k=d_{12}=0$, and introduce the rotated mode operators $b_k$
as
\begin{eqnarray}a_1=b_1\cos(\phi)+b_2\sin(\phi), \quad
a_2=b_2\cos(\phi)-b_1\sin(\phi), \label{rotation}
\end{eqnarray}
where $\tan(2\phi)=2\gamma_{12}/(\gamma_2-\gamma_1)$. Then for the
non-unitary part of the master equation (\ref{master_markov_term})
one has
\begin{eqnarray}
\nonumber (\gamma_1-\gamma_{12})\mathcal{L}(a_1)+
(\gamma_2-\gamma_{12})\mathcal{L}(a_2)+\gamma_{12}\mathcal{L}(a_1+a_2)
\longrightarrow {\bar \gamma}_1\mathcal{L}(b_1)+
{\bar\gamma}\mathcal{L}(b_2),
\end{eqnarray}
where
${\bar\gamma}_1=\gamma_1\cos^2(\phi)+\gamma_2\sin^2(\phi)-\gamma_{12}\sin(2\phi)$,
${\bar\gamma}_2=\gamma_2\cos^2(\phi)+\gamma_1\sin^2(\phi)+\gamma_{12}\sin(2\phi)$.
Obviously, if the transformation (\ref{rotation}) leaves the form
of the Hamiltonian (\ref{h_0_kerr}) invariant, one can derive an
exact solution in the way described in this Section.


\section{Uncorrelated reservoirs}

\subsection{General solution for the uncorrelated reservoirs}

In this Section we consider specific effects of the uncorrelated losses
in the process of the cross-Kerr nonlinear interaction. Notably, the losses in such a nonlinear
process can lead to loss-mediated correlations between the modes, as seen from Eq.~(\ref{ham}).
These intermodal correlations modify the effect of losses on the quantum state generated in the cross-Kerr interaction with respect to the result, which one would intuitively expect treating the generation and
loss separately, i.e. subjecting to loss a state that has been produced without losses.

Consider a particular problem of generating an entangled two-mode
state from initially uncorrelated coherent states. We assume that
the nonlinearity is given purely by the cross-Kerr interaction,
i.e. we put $\chi_{kl}=(1-\delta_{kl})\chi/2$ in the Hamiltonian
(\ref{h_0_kerr}). Producing entangled states this way is important
in a number of schemes of quantum computation and communication
using continuous variables
\cite{{Tyc08},{clancy2008},{spiller2005}}. We assume modes $a_1$
and $a_2$ to be initially in coherent states with amplitudes
$\alpha_1$ and $\alpha_2$, respectively. As was pointed in the
previous Section, for this choice of initial states, the solution
given by Eq.~(\ref{ham}) has a simple form:
\begin{eqnarray} \nonumber
|\rho(t)\rangle
=\exp\left[-\frac{[d_1(A_0^{(1)})^2+d_2(A_0^{(2)})^2]t}2\right]\exp\left\{\ii\chi t
  (a_1^\dagger a_1a_2^\dagger a_2-\tilde a_1^\dagger\tilde a_1\tilde a_2^\dagger\tilde a_2)
   -\frac{\gamma_1t}2(a_1^\dagger a_1+\tilde a_1^\dagger\tilde a_1)
   -\frac{\gamma_2t}2(a_2^\dagger a_2+\tilde a_2^\dagger\tilde a_2)\right\}\\  \times
\exp\left[\frac{\gamma_1\left(\e^{\ii\chi t(a_2^\dagger a_2-\tilde
  a_2^\dagger\tilde a_2)-\gamma_1t}-1\right)}{\ii\chi (a_2^\dagger a_2-\tilde
  a_2^\dagger\tilde a_2)-\gamma_1}\,|\alpha_1|^2\right]
\exp\left[\frac{\gamma_2\left(\e^{\ii\chi t(a_1^\dagger a_1-\tilde
  a_1^\dagger\tilde a_1)-\gamma_2t}-1\right)}{\ii\chi (a_1^\dagger a_1-\tilde
  a_1^\dagger\tilde a_1)-\gamma_2}|\alpha_2|^2\right]
  |\alpha_1\rangle|\tilde\alpha_1^*\rangle|\alpha_2\rangle|\tilde\alpha_2^*\rangle.
  \quad
\label{evolution}
\end{eqnarray}
A distinctive feature of the solution (\ref{evolution}) is that
the exponential terms contain only operators diagonal in the
number-state basis. Thus, in this basis the solution
(\ref{evolution}) turns into
\begin{eqnarray} \nonumber
|\rho(t)\rangle&=&\e^{-|\alpha_1|^2-|\alpha_2|^2}\sum\limits_{k,l,m,n=0}^{\infty}
\frac{\alpha_1^k\alpha_1^{*l}\alpha_2^m\alpha_2^{*n}}{\sqrt{k!l!m!n!}}
\exp{\left\{-\frac{1}{2}[d_1(k-l)^2+d_2(m-n)^2]t\right\}} \\
&&\times\exp{\left\{\ii\chi t (km-ln)
-\frac{\gamma_1t}{2}(k+l)-\frac{\gamma_2t}2(m+n)+f^{(2)}_{mn}(t)+f^{(1)}_{kl}(t)\right\}}
|k\rangle|\tilde l\rangle|m\rangle|\tilde n\rangle,
\label{evolution_fock}
\end{eqnarray}
where
\begin{eqnarray} 
f^{(2)}_{mn}(t)=\frac{\gamma_1(\e^{\ii \chi
t(m-n)-\gamma_1t}-1)}{\ii \chi (m-n)-\gamma_1}\,|\alpha_1|^2,\quad
f^{(1)}_{kl}(t)= \frac{\gamma_2(\e^{\ii\chi
t(k-l)-\gamma_2t}-1)}{\ii\chi (k-l)-\gamma_2}|\alpha_2|^2.
\label{f-functions}
\end{eqnarray}


\subsection{Analysis: When is the purity of the state not broken by damping?}

Now let us analyze the solution (\ref{evolution_fock}) in more
detail and consider for the moment the case of no dephasing
($d_1=d_2=0$). In this case  the expression for the purity of the
state given by the solution (\ref{evolution_fock}) is quite
similar in structure to this solution itself:
\begin{eqnarray} \nonumber
{\rm
Tr}\{(\rho(t))^2\}&=&\e^{-2|\alpha_1|^2-2|\alpha_2|^2}\sum\limits_{k,l,m,n=0}^{\infty}
\frac{|\alpha_1|^{2(k+l)}|\alpha_2|^{2(m+n)}}{{k!l!m!n!}}
\\
&&\times\exp{\left\{-{\gamma_1t}(k+l)-{\gamma_2t}(m+n)+2{\rm
Re}f^{(2)}_{mn}(t)+2{\rm Re}f^{(1)}_{kl}(t)\right\}}.
\label{evolution_purity}
\end{eqnarray}

To describe effects of simultaneous damping (photon loss) and
cross-Kerr nonlinearity on the generated state, consider first a
`short-time regime' where the following relations hold for all the
photon numbers $k,l$ ($m,n$) that have a non-negligible
probability of occurring in the state $|\alpha_1\rangle$
($|\alpha_2\rangle$):
\begin{eqnarray}
|\ii\chi t(m-n)-\gamma_1t| < 1, \quad |\ii \chi t(k-l)-\gamma_2t| < 1.
\label{small time conditions}
\end{eqnarray}
Then expressions (\ref{f-functions}) can be expanded  as
\begin{eqnarray}
f^{(2)}_{mn}(t)=\gamma_1t|\alpha_1|^2\sum\limits_{j=0}^{\infty}{[\ii \chi
t(m-n)-\gamma_1t]^j\over(j+1)!}, \quad
f^{(1)}_{kl}(t)=
\gamma_2t|\alpha_2|^2\sum\limits_{j=0}^{\infty}{[\ii \chi
t(k-l)-\gamma_2t]^j\over(j+1)!}. \label{f-functions_expand}
\end{eqnarray}
The first three terms of these expansions already describe quite well
typical effects produced by simultaneous damping and cross-Kerr
nonlinearity. For example, for $f^{(2)}_{mn}(t)$ in the limit of
small times one has
\[f^{(2)}_{mn}(t) \approx \gamma_1t|\alpha_1|^2\left(1-{\gamma_1t\over2}+\ii {\chi t\over2}(m-n)
+{1\over6}[\gamma_1^2t^2-\chi^2t^2(m-n)^2]-{\ii \over6}\gamma_1\chi
t^2(m-n)\right).
\]
Obviously, first three terms in the round brackets of this
expressions do not lead to breaking of the purity of the state.
Retaining only them in expansion renders unity value for the
right-hand side of Eq.(\ref{evolution_purity}). Indeed, assuming
\begin{eqnarray}1-{\gamma_1t\over2} \gg
{1\over6}|\gamma_1^2t^2-\chi^2t^2(m-n)^2|, \quad
1-{\gamma_2t\over2} \gg {1\over6}|\gamma_2^2t^2-\chi^2t^2(k-l)^2|,
\label{f-functions_conditions_small}
\end{eqnarray}
one obtains the time-dependent density matrix
formally coinciding with the result for no photon
loss \cite{Tyc08}:
\begin{eqnarray}
|\rho(t)\rangle
\approx\exp\{-|\alpha_1(t)|^2\}\sum_{k,l=0}^\infty
\frac{[\alpha_1(t)]^k[\alpha_1^*(t)]^l}{\sqrt{k!l!}}
|k\rangle|\tilde l\rangle|\alpha_2(t)\e^{\ii\chi k t}\rangle
\overline{|\alpha_2^*(t)\e^{\ii\chi lt}}\rangle.
\label{rho_no_dephasing}
\end{eqnarray}
Here time-dependent amplitudes do not depend on the numbers
$k,l$:
\begin{eqnarray}
\alpha_1(t)=\alpha_1\exp\left\{-\frac{\gamma_1t}{2}-{\ii \over2}\gamma_2\chi|\alpha_2|^2t^2\right\},
\quad
\alpha_2(t)=\alpha_2\exp\left\{-\frac{\gamma_2t}{2}-{\ii \over2}\gamma_1\chi|\alpha_1|^2t^2\right\}.
\label{functions1}
\end{eqnarray}
The state given by Eqs. (\ref{evolution_fock}) remains negligibly
affected by losses if $\gamma_{1,2}t\ll1$. Thus, the
considered scheme of non-classical state generation is quite
robust with respect to photon loss (in drastic difference with
propagation losses of already generated cat-state where the
off-diagonal terms  $|k\rangle\langle l|, k\not=l$ will decay with
the rates proportional to $\gamma |\alpha|^2$). In addition, it is interesting to note,
that one might be able to satisfy conditions
$\gamma_{1,2}t\ll1$ in schemes involving dispersive atom-field
interactions in QED where it is possible to restrict losses
to the photon loss of cavity modes \cite{plenio08}. Moreover,
further in this work we consider ways to circumvent an influence
of losses by making them correlated.

Remarkably, a purity of the generated bimodal state can be preserved
not only in the case of small losses, but also for large loss.
Indeed, in the limits of large
losses one can consider $\chi (k-l)$ and $\chi (m-n)$ as small
quantities and expand functions (\ref{f-functions})  in the
following manner:
\begin{eqnarray}
|\alpha_2|^{-2}f^{(1)}_{kl}(t)\approx
(1-\e^{-\gamma_2t})-\ii \chi(k-l)\left[\left(t+{1\over\gamma_2}\right)\e^{-\gamma_2t}-{1\over\gamma_2}\right]+
\chi^2(k-l)^2\left[\left({t^2\over2}+{t\over\gamma_2}+{1\over\gamma_2^2}\right)\e^{-\gamma_2t}-
{1\over\gamma_2^2}\right], \\
\nonumber |\alpha_1|^{-2}f^{(2)}_{mn}(t)\approx
(1-\e^{-\gamma_1t})-\ii \chi(m-n)\left[\left(t+{1\over\gamma_1}\right)\e^{-\gamma_1t}-{1\over\gamma_1}\right]+
\chi^2(m-n)^2\left[\left({t^2\over2}+{t\over\gamma_1}+{1\over\gamma_1^2}\right)\e^{-\gamma_1t}-
{1\over\gamma_1^2}\right]. \label{tailor}
\end{eqnarray}
Note that this approximation holds for arbitrary interaction
times. If the interaction time is sufficiently large to fulfill the
conditions
\begin{eqnarray}
1-\e^{-\gamma_2t} \gg
\chi^2(k-l)^2\left|\left({t^2\over2}+{t\over\gamma_2}+{1\over\gamma_2^2}\right)\e^{-\gamma_2t}-
{1\over\gamma_2^2}\right|, \\
1-\e^{-\gamma_1t} \gg
\chi^2(m-n)^2\left|\left({t^2\over2}+{t\over\gamma_1}+{1\over\gamma_1^2}\right)\e^{-\gamma_1t}-
{1\over\gamma_1^2}\right|, \label{long_time_conditions}
\end{eqnarray}
then the state given by Eqs. (\ref{evolution_fock}) is practically
pure. Under the conditions (\ref{long_time_conditions}) this state
is of the form described by Eq.~(\ref{rho_no_dephasing}) with the
time-dependent amplitudes given by
\begin{eqnarray}
\nonumber
\alpha_1(t)=\alpha_1\exp\left\{-\frac{\gamma_1t}{2}-i\chi|\alpha_2|^2
\left[t\e^{-\gamma_2t}-\gamma_2^{-1}(1-\e^{-\gamma_2t})\right]\right\},
\\
\alpha_2(t)=\alpha_2\exp\left\{-\frac{\gamma_2t}{2}-\ii\chi|\alpha_1|^2
\left[t\e^{-\gamma_1t}-\gamma_1^{-1}(1-\e^{-\gamma_1t})\right]\right\}.
\label{functions}
\end{eqnarray}
Clearly, purity of the resulting state in the long time-limit
is precisely a consequence of a strong photon loss. In this way strong photon
loss paradoxically suppresses state mixing predicted by the general solution
(\ref{evolution_fock}).


\subsection{Survival of non-classicality for large losses}

There is another interesting feature that distinguishes the losses occurring in the process of
Kerr interaction from the losses that take place after the interaction.  In particular, the non-Gaussian state
generated by the cross-Kerr interaction in the scheme discussed in
\cite{Tyc08} can retain its non-classical features even for the
loss level, which would completely eliminate any such features in case
of free propagation of the state.  To be specific, a typical signature
of non-classicality of a quantum state is the fact that its Wigner
function  is negative in some regions of the phase space
\cite{milburn06}. We will show that for a 50\% photon loss
occurring in the scheme \cite{Tyc08} during the cross-Kerr
interaction, the Wigner function of the output state retains its
negativity while the same photon loss occurring after the
interaction would make the Wigner function necessarily positive.
\begin{figure}[h]
\includegraphics[width=10cm]{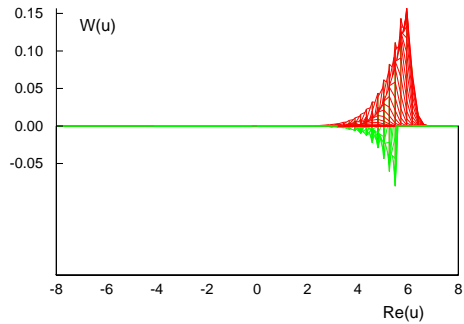}
\caption{(Color online) Wigner function $W(u)$ of the
non-classical non-Gaussian state of Ref.~\cite{Tyc08} exhibits
strong negativity. Wigner function is viewed along the imaginary
axis of the $u$ plane. } \label{wigner0}
\end{figure}

\begin{figure}[h]
\includegraphics[width=10cm]{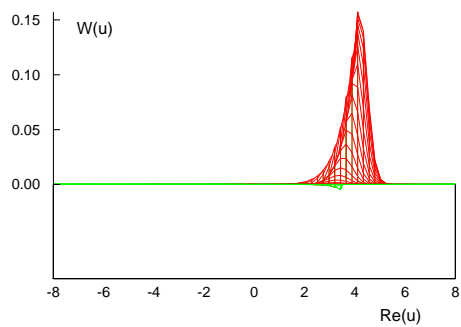}
\caption{(Color online) Wigner function $W(u)$ for about $50\%$
photon loss in mode $a$ corresponding to $\gamma_at=0.7$. The
negative region of the Wigner function is now negligible, but
still present, and the amplitude of the state is clearly damped. }
\label{wigner07}
\end{figure}

In the scheme \cite{Tyc08} for generating non-classical states, two coherent
states in modes $a_1,a_2$ ($a,b$ in notations of \cite{Tyc08})
interact via cross-Kerr effect in a non-linear medium and
subsequently the $x$-quadrature of mode $a_2$ is measured. The resulting state
of mode $a_1$ exhibits Wigner function with negative regions (see
Fig.~\ref{wigner0}) and a characteristic crescent (or banana) shape.  Suppose
mode $a_1$ is subject to losses during the cross-Kerr interaction, i.e.,
$\gamma_1>0$ in Eq.~(\ref{evolution}) while we assume $\gamma_2=d_1=d_2=0$.
The photon loss in mode $a_1$ is given by the reduction of the coherent
amplitude described by Eq.~(\ref{functions1}). Consider the situation when the
mean photon loss is $50\%$. This corresponds to ${\rm
e}^{-\gamma_1t}=1/2\,\Rightarrow\,\gamma_1t=\ln2\approx0.69$.  The plot of the
corresponding Wigner function is shown in Fig.~\ref{wigner07} which shows
clearly that although the negative region of the Wigner function is surpressed,
it is still present.

Now compare this with the situation when losses are introduced to mode $a_1$
after the lossless cross-Kerr interaction has taken place. Such losses are
equivalent to mixing mode $a_1$ with the vacuum state on a beam splitter (BS)
and discarding one BS output.
A 50\% loss corresponds to a 50/50 BS. It is known~\cite{Kim95} that for such a
balanced beam splitter the Wigner function of one BS output
can be expressed as a scaled Husimi Q-function of the input state:
\begin{equation}
  W_{\rm out}(\alpha)=2Q_{\rm in}(\sqrt2\,\alpha).
\end{equation}
The Q-function of a state $\rho$ is defined as
$Q(\alpha)=\langle\alpha|\rho|\alpha\rangle/\pi$ and is clearly non-negative
for all coherent state amplitudes $\alpha$, i.e., in the whole phase space of
the mode. Therefore negative regions of the Wigner function cannot survive
losses larger than 50\% if these occur during propagation of the generated
state.
Hence losses that take place in the process of state generation via the cross-Kerr interaction are less
harmful to the non-classicality of the output state than the same level of loss after the interaction.


\subsection{Dephasing into independent reservoirs}

As can be seen from the solution (\ref{evolution_fock}), an
influence of dephasing into independent reservoirs can be
profoundly destructive. Dephasing leads to diminishing of the
off-diagonal elements in the number-state basis with rates
proportional to the difference of these numbers. So for the
coherent state with the amplitude $\alpha_k$, a
condition $d_k|\alpha_k|^4t \ll 1$ should be fulfilled for the interaction time $t$
to consider the influence of dephasing negligible. In the
recently discussed QED schemes  for
generating non-linearity (including EIT-like ones), dephasing is usually disregarded without
being estimated (see, for example, \cite{plenio08}). However, since
the rate of losses increases with increasing intensity of the
coherent states used in the generation process, more caution is required with respect to the dephasing. For schemes
involving large cross-Kerr nonlinearity in the solid-state
structures (such as, for example, photonic crystals)
emitter-mediated dephasing could be a major source of
state decoherence and a reason for failure of schemes involving
initial coherent states with large number of photons. However,  remarkably, if the cross-Kerr
nonlinear interaction scheme is designed in such a way that losses due to dephasing
are correlated, then it may be possible to avoid their destructive
effects. This is the subject of the next section.


\section{Correlated reservoirs}

It is well established, that the states of quantum systems can be
correlated and even entangled through interaction with the common
reservoir \cite{{braun},{horhammer}}. This phenomenon can occur
even in absence of any direct interaction between systems. In in
Appendix B3 we give an example of such a phenomenon for a scheme
of generating the cross-Kerr nonlinearity via dispersive
interaction of modes with emitters. There a beam-splitting action
of the common reservoir is considered, and it is shown how such a
reservoir can produce a stationary entangled state of two modes.

Here we focus our attention on another important possibility:
namely, on a way to neutralize a destructive influence of losses
in the process of generation by rendering these losses correlated
and exploit a correlating effect of the reservoir.


\begin{figure}[th]
\centerline{\psfig{file=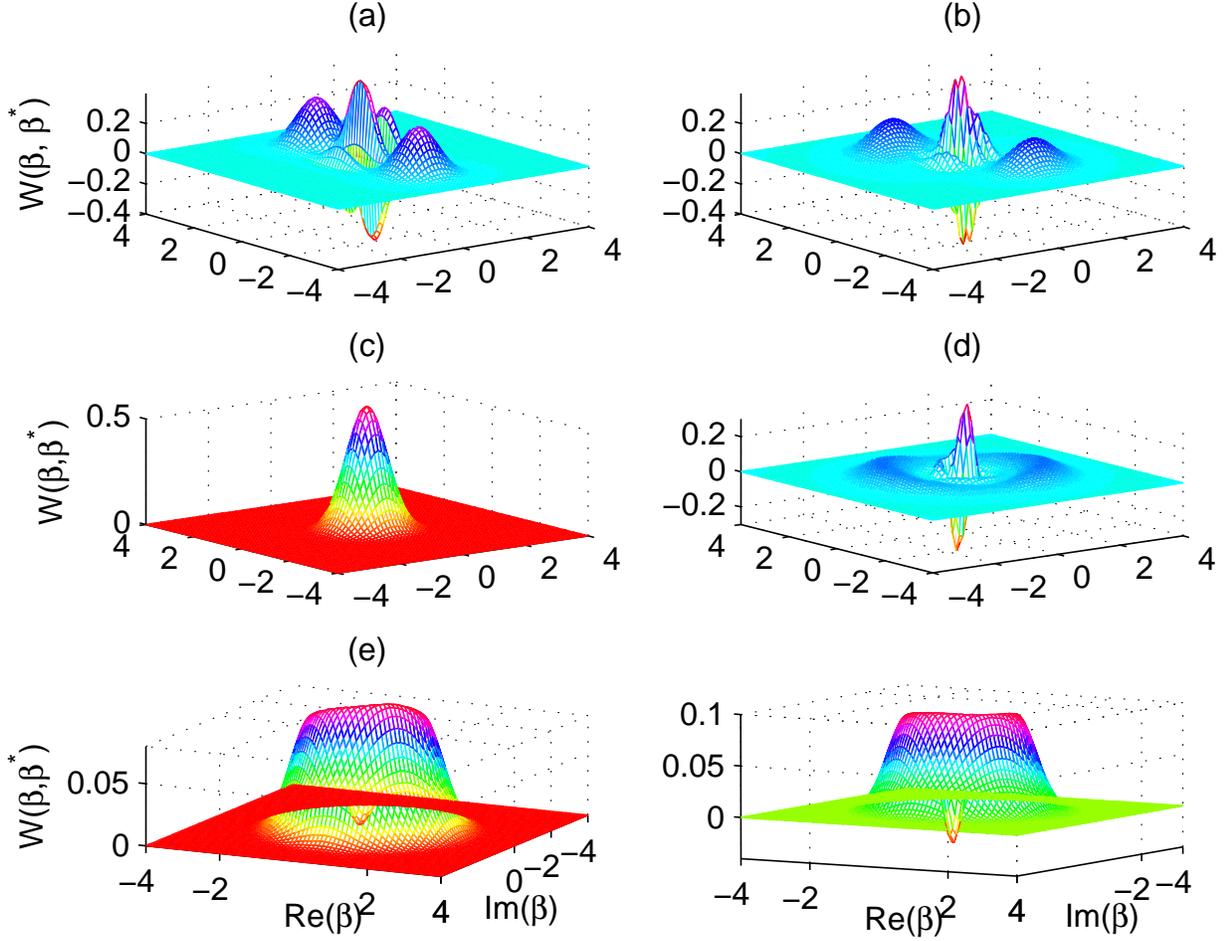,width=\linewidth}} \vspace*{8pt}
\caption{(Color online) Examples of the Wigner function for the
conditioned cat state of the rotated mode $b_1$. Figure (a)
corresponds to absence of loss. Figure (b) corresponds to the
perfectly correlated loss, $\gamma_1=\gamma_2=\gamma_{12}=10\chi$;
for figure (d) $\gamma_1=\gamma_2=\gamma_{12}=3\chi$. Figure (c)
corresponds to the completely uncorrelated loss,
$\gamma_1=\gamma_2=3\chi$, $\gamma_{12}=0$. Figure (f) corresponds
to the partially correlated loss, $\gamma_1=\gamma_2=3\chi$,
$\gamma_{12}=2.95\chi$; figure (e) corresponds to the completely
uncorrelated loss $\gamma_1=\gamma_2=0.5\chi$, $\gamma_{12}=0$.
For all figures $\chi t=\pi/2$. } \label{fig2}
\end{figure}


Let us consider an example of the realistic scheme to produce the
large cross-Kerr nonlinearity described in Section~\ref{lambda-system} (see Eq.~(\ref{hamiltonian_lambda_system}) and the text thereafter). We
consider the case of
\[{g_1^2\over\Delta_1-\delta}={g_2^2\over\Delta_1+\delta}=-\chi{2\Omega^2\over \Delta_2},\]
so the master equation (\ref{master_lambda_model}) now transforms as
\begin{eqnarray}
{\dd\over \dd t}\rho(t)\approx {\ii \chi}\left[\left(
a_1^{\dagger}a_1+
a_2^{\dagger}a_2\right)^2,\rho(t)\right]+{1\over2}\left[(\gamma_1-\gamma_{12})
\mathcal{L}(a_1)+(\gamma_2-\gamma_{12})\mathcal{L}(a_2)\right]\rho(t)+
{\gamma_{12}\over2}\mathcal{L}(a_1+a_2)\rho(t).
\label{master_lambda_model1}
\end{eqnarray}
In absence of decoherence (i.e., $\gamma_k=\gamma_{12}=0$), the
scheme described by Eq.~(\ref{master_lambda_model}) is able to
generate entangled superpositions of Schr\"odinger-cat states from
initial coherent states of modes $a_1$ and $a_2$ (for similar
schemes see, for example, \cite{{yurke},{mecozzi}}). It is easy to
see that for $\chi t=\pi/2$ the scheme produces an entangled
superposition of coherent states from a pair of initially
uncorrelated coherent states, which reads as
\begin{eqnarray}
\exp\left\{\ii{\pi\over2}\left(a_1^{\dagger}a_1+
a_2^{\dagger}a_2\right)^2\right\}|\alpha_1\rangle|\alpha_2\rangle
={1\over\sqrt{2}}\left(\ii |\alpha_1\rangle|\alpha_2\rangle+|-\alpha_1\rangle|-\alpha_2\rangle\right).
\label{lambda_nodissipation_exact}
\end{eqnarray}
In the presence of correlated reservoirs one can find a solution of
Eq.~(\ref{master_lambda_model1}) noticing that under the rotation
(\ref{rotation}) the Hamiltonian part of Eq.
(\ref{master_lambda_model1}) remains invariant. Thus, performing
the rotation one obtains
\begin{eqnarray}
{\dd\over \dd t}\rho(t)\approx {\ii \chi}\left[\left(
b_1^{\dagger}b_1+
b_2^{\dagger}b_2\right)^2,\rho(t)\right]+{1\over2}\left[{\bar\gamma}_1
\mathcal{L}(b_1)+{\bar\gamma}_2\mathcal{L}(b_2)\right]\rho(t)
\label{master_lambda_model2}
\end{eqnarray}
with the exact solution given by Eq.~(\ref{ham}). Also, one
immediately sees that for the completely correlated reservoirs
(i.e. $\gamma_1\gamma_2=\gamma_{12}^2$),
the mode $b_1$ is not affected by the
loss, as ${\bar{\gamma}_1}=0$ then. Naturally,  a cat state can be generated in this mode. Thus,
in the limit of large loss, ${\bar\gamma}_2t\gg 1$,  and for $\chi
t =\pi/2$ it follows from Eq.~(\ref{master_lambda_model2}) that
\begin{eqnarray}
|\rho(\pi/2\chi)\rangle\approx|\Psi\rangle|{\overline{\Psi}}\rangle;
\quad |\Psi\rangle={1\over\sqrt{2}}\left(\ii
|\tilde{{\alpha}}_1\rangle|\tilde{\alpha}_2\rangle+|
-\tilde{\alpha}_1\rangle|-\tilde{\alpha}_2\rangle\right),
\label{lambda_dissipation_exact}
\end{eqnarray}
where
${\tilde{\alpha}}_1=\alpha_1\cos^2(\phi)-\alpha_2\cos(\phi)\sin(\phi)$,
${\tilde{\alpha}}_2=\alpha_2\sin^2(\phi)-\alpha_1\cos(\phi)\sin(\phi)$.
We also assume that
$2\chi|\alpha_1\cos(\phi)-\alpha_2\sin(\phi)|^2\ll{\bar\gamma}_2$.
From Eq.~(\ref{lambda_dissipation_exact}), we derive a
conclusion that the only effect of completely correlated loss is a
reduction of amplitudes of the coherent states forming the
superposition (\ref{lambda_nodissipation_exact}).

Thus, we have seen that by making losses completely correlated one
can completely avoid decoherence caused by these losses. Of
course, in practice one can hardly have completely
correlated reservoirs  due to presence of additional uncorrelated loss (such
as modal losses due to coupling to additional reservoirs etc.). Nevertheless,
designing the
scheme such as to have predominantly correlated losses might greatly
enhance its robustness in production of non-classical states. We illustrate this
with the simple example of the conditioned cat-state generation
from the solution of Eq.~(\ref{master_lambda_model}). If the
rotated mode $b_2$ impinges on the detector, in the case of no
signal on the detector the rotated mode $b_1$
is (up to
the normalization factor) in the state:
\begin{eqnarray}
|\rho_1(t)\rangle \sim \sum\limits_{k,l=0}^{\infty}
\frac{{\bar\alpha}_1^k{\bar\alpha}_1^{*l}}{\sqrt{k!l!}}
\exp{\left\{\ii\chi t (k^2-l^2)
-\frac{{\bar\gamma}_1t}{2}(k+l)+f_{kl}(t)\right\}}
|k\rangle|\tilde l\rangle , \label{evolution_fock2}
\end{eqnarray}
where ${\bar\alpha}_1=\alpha_1\cos(\phi)-\alpha_2\sin(\phi)$,
${\bar\alpha}_2=\alpha_2\cos(\phi)+\alpha_1\sin(\phi)$, and
\[f_{kl}(t)=\frac{{\bar\gamma}_1(\e^{\ii \chi
t(k-l)-{\bar\gamma}_1t}-1)}{\ii \chi
(k-l)-{\bar\gamma}_1}|{\bar\alpha}_1|^2 +
\frac{{\bar\gamma}_2(\e^{\ii\chi
t(k-l)-{\bar\gamma}_2t}-1)}{\ii\chi
(k-l)-{\bar\gamma}_2}|{\bar\alpha}_2|^2.
\]

In Fig.~\ref{fig2} one can see examples of the Wigner function of
the state (\ref{evolution_fock2}). For no loss
(Fig.~\ref{fig2}(a)) and $\chi t=\pi/2$ the state
(\ref{evolution_fock2}) is a usual Schr\"odinger cat state with
the pronounced oscillations near the origin. Large correlated loss
($\gamma_{1,2} \gg \chi$) changes the size of the cat and rotates
it (Fig.~\ref{fig2}(b)), but otherwise leaves it intact. Lower
correlated loss distorts the cat (Fig.~\ref{fig2}(d)) due to
influence of additional mixing between modes in the interaction
process (as it follows from Eq.~(\ref{evolution_fock2})).
Nevertheless, the state is strongly non-classical. Uncorrelated
loss with the same rate eliminates the non-classicality outright
(Fig.~\ref{fig2}(c)). Correlated loss allows for the
non-classicality to survive (Fig.~\ref{fig2}(f)) even if the
uncorrelated loss with the rate equal to difference between
individual rates and the correlation rate (e.g.,
$\gamma=\gamma_1-\gamma_{12}$) destroys the non-classicality
completely (Fig.~\ref{fig2}(e)).


\section{Conclusions}

In what presented here, we followed the quest to find ways to impair the decoherence processes in quantum state generation and manipulation. For the particular class of nonlinear interaction processes, we have found two striking examples of loss dynamics, for which the losses themselves counteract decoherence: loss-mediated correlations between the interacting modes and losses to the correlated reservoirs. The latter result in strongly correlated modal loss.

These correlated losses influence the dynamics of the modes undergoing
the cross-Kerr nonlinear interaction
in a completely different way than losses
into independent reservoirs, the aspect of quantum nonlinear dynamics to large extent unexplored so far.
Thus, remarkably, designing the schemes for the generation of the
Kerr nonlinearity  in such a way that losses
in this process are correlated, one can
greatly diminish their destructive impact and even exploit them for entanglement generation (see also Appendix B).
As to the origin of this effect, if both modes, for example, interact with the same emitter transition,
emitter losses are likely to lead to the correlated loss
of both of these modes. Note,
that in this case a significant modal loss might occur even in the case when
the emitter subject to losses stays in superposition of
metastable levels with negligibly small probability to occupy
higher, decaying levels. For more details and examples on the origin of the correlated losses and their entangling effect see Appendix B.

Turning to the other aforementioned unexpected aspect of the quantum dynamics, we have demonstrated that losses in the nonlinear process of state generation affect the quantum state in quite a different
way to the propagation losses of already generated nonclassical
state.  This is mainly due to the loss-mediated correlations between the modes participating in the cross-Kerr interaction. In addition, losses through coupling to the correlated reservoirs (correlated loss) further enhance the difference between the decoherence processes {\it during} and {\it after} the state generation. In particular, non-classicality seems to be more
robust with respect to the generation loss than to the propagation
loss. We discussed an example of generation loss exceeding $50\%$ with negative values of the Wigner function
preserved, whereas the
propagation loss exceeding $50\%$ renders the Wigner function
completely positive.

\section*{Acknowledgments}

The authors  gratefully acknowledge funding from the EU project
FP7-212008 COMPAS (N. K. and T.T.),  EQUIND project of 6FP
IST-034368, FFR of Belarus project F08P-131 and FAPESP, Brazil,
project 2008/57657-3 (D.M.). Also, D.M. would like to thank the
School of Physics and Astronomy, University of St. Andrews, for
their hospitality.

\appendix
\section{Derivation of the master equation}

We start from the general effective Hamiltonian $H(t)$ of Eq.~(\ref{ham_ger}, \ref{vloss})
describing both self- and cross-interaction  and
interaction of modes with reservoirs responsible for losses (see Section~\ref{master_section}):
\begin{eqnarray}
V(t)=H(t)+V_{\rm loss}(t), \qquad
V_{\rm loss}(t)=a_1^{\dagger}\Gamma_1(t)+\Gamma_1^{\dagger}(t)a_1+a_2^{\dagger}\Gamma_2(t)+\Gamma_2^{\dagger}(t)a_2+
a_1^{\dagger}a_1 D_1(t)+a_2^{\dagger}a_2D_2(t).\nonumber
\end{eqnarray}
Setting $\hbar= 1$ for sake of simplicity, and using a time-convolutionless
projection operator technique, one obtains the following master
equation \cite{breuer}:
\begin{eqnarray}
\nonumber {\dd\over \dd t}\rho(t)&=&-\ii[H(t),\rho(t)] - \\
&&\int\limits_0^{t}\dd\tau \left\{\langle
V_{\rm loss}(t)V_{\rm loss}(\tau)\rangle_{\rm
r}\rho(t)+\rho(t)\langle V_{\rm loss}(\tau)V_{\rm
loss}(t)\rangle_{\rm r} - \langle V_{\rm loss}(\tau)\rho(t)V_{\rm
loss}(t)\rangle_{\rm r}- \langle V_{\rm loss}(t)\rho(t)V_{\rm
loss}(\tau)\rangle_{\rm r}\right\}\qquad \label{master_gener}
\end{eqnarray}
where $\rho(t)$ denotes the density matrix of the system averaged
over states of all reservoirs (and over possible stochastic
variables, too), and $\langle \ldots \rangle_{\rm r}$ denotes
averaging over all reservoirs.
Reservoir modes are assumed to be initially in the vacuum states.
Further, we consider reservoirs of different types
to be independent; correlation functions
$\langle \Gamma_{j}(t)D_{k}^{\dagger}(\tau) \rangle_{\rm r}$, $j,k=1,2$
are taken to be zero. We consider the reservoirs of the same kind to be
mutually correlated, i.e., for Markovian dephasing reservoirs we
assume
\begin{eqnarray}
\nonumber \langle D_{i}(t)D_{i}(\tau) \rangle_{\rm r}={1\over 2
}d_{i}\,\delta(t-\tau), \quad \langle D_{1}(t)D_{2}(\tau)
\rangle_{\rm r}={1\over 2 }d_{12}\,\delta(t-\tau).
\end{eqnarray}
Here $\delta(t-\tau)$ is the delta-function and the rates $d_{1}, d_2$, are
real and non-negative. The cross-correlation parameter $d_{12}$
is taken to be real. For the photon loss reservoir, we assume that
the term
$a_1^{\dagger}\Gamma_1(t)+\Gamma_1^{\dagger}(t)a_1+a_2^{\dagger}\Gamma_2(t)+\Gamma_2^{\dagger}(t)a_2$
in the  Hamiltonian (\ref{ham_ger}) preserves the total number of
photons, i.e. only non-zero correlation functions are
\begin{eqnarray}
 \langle \Gamma_{i}(t)\Gamma_{i}^{\dagger}(\tau)
\rangle_{\rm r}={1\over 2 }\gamma_{i}\,\delta(t-\tau), \quad
\langle\Gamma_{1}(t)\Gamma_{2}^{\dagger}(\tau) \rangle_{\rm
r}={1\over 2 }\gamma_{12}\,\delta(t-\tau).
\label{gamma12}
\end{eqnarray}
Also here, for simplicity, the
rates $\gamma_{1}, \gamma_2$  are assumed to be real and non-negative and
the cross-correlation parameter $\gamma_{12}$ to be
real. For self-loss rates, $\gamma_{i}$, $d_i$, and `cross'-loss
rates, $\gamma_{12}$, $d_{12}$, the following relations hold
\begin{eqnarray}d_{a}d_{b}\geq d_{ab}^2, \quad \gamma_{a}\gamma_{b}\geq
\gamma_{ab}^2. \label{rates_unequalities}
\end{eqnarray}

Under the assumptions made above, the  master equation of Eq.~(\ref{master_markov}) is obtained from Eq.~(\ref{master_gener}).

\section{Origin of correlated loss}

\subsection{An example: dispersive two-mode Jaynes-Cummings model
with damping}

We will illustrate the process of appearance of a {\it correlated
modal photon loss} in a process of off-resonant interaction
between a mode and an emitter with a simple example. Consider two
modes of the same frequency interacting off-resonantly with just a
single two-level system (TLS), see Fig.~\ref{fig1}(a).  In the
rotating-wave approximation and in the interaction picture with
respect to the free Hamiltonian of the reservoir, in the frame
rotating with the TLS transition frequency $\omega_0$, one has the
following Hamiltonian describing the problem:
\begin{eqnarray}
H_{JK}=  \Delta (a_1^{\dagger}a_1+a_2^{\dagger}a_2)+
\left(\sigma^+(g_1a_1+g_2a_2)+(g_1a_1^{\dagger}+g_2a_2^{\dagger})\sigma^-\right)+
\left(\sigma^+\Gamma(t)+\Gamma^+(t)\sigma^-\right).
\label{jkmodel_ham_spon}
\end{eqnarray}
Here $\Delta=\omega_0-w$ and $w$ is the mode frequency; $g_{1,2}$
are interaction constants for the corresponding modes;
$\sigma^{\pm}$ and $\sigma_z$ are Pauli operators for the TLS,
$\sigma^{+}=|2\rangle\langle 1|$, $\sigma^{-}=|1\rangle\langle
2|$, $\sigma_z=\sigma^+\sigma^--\sigma^-\sigma^+$;  vectors
$|k\rangle$, $k=1,2$ describe the lower and the
upper TLS levels, correspondingly. The reservoir operator $\Gamma(t)$ describes
the TLS energy loss. We assume this reservoir to be
Markovian  and the following relations hold:
\[ \langle\Gamma(t)\rangle_{\rm r}=0, \quad
\langle\Gamma(t)\Gamma^{\dagger}(\tau)\rangle_{\rm r}=\gamma\,
\delta(t-\tau)
\]
where brackets $\langle\ldots\rangle_{\rm r}$ denote an averaging
over the reservoir. Here we are assuming that losses are weak
(i.e., $\gamma \ll |g_{1,2}|$).

We adopt the usual conditions for an adiabatic
elimination of the emitter, i.e. the TLS starts at the lower
level, and the detuning  $\Delta$ between the mode frequency and the TLS
transition frequency is much larger than $g_{1,2}$.
Thus,  the TLS upper level remains practically unpopulated. Changing to
the interaction picture with respect to the part of the
Hamiltonian (\ref{jkmodel_ham_spon}) corresponding to the absence
of the TLS-field interaction, we get the following interaction
Hamiltonian
\begin{eqnarray}
V_{JK}(t)=  G \left(\sigma^+C \exp\{\ii\Delta t \}+h.c\right)+
\left(\sigma^+\Gamma(t)+\mbox{H. c.}\right) \label{jkmodel_ham_V_spon}
\end{eqnarray}
with the bosonic annihilation operator for the collective mode
\[C={1\over \sqrt{g_1^2+g_2^2}}\,(g_1a_1+g_2a_2)
\]
and $G=\sqrt{g_1^2+g_2^2}$. A formal solution for $\sigma^+(t)$
without losses up to the third-order terms can be approximated as
\begin{eqnarray}
\sigma^+(t)\approx\sigma^+(0)+\ii\int\limits_{0}^t\dd t_1
[2F^{\dagger}(t_1)F(t_1)-1]X^{\dagger}(t_1),
\label{jkmodel_solution_sigma_spon}
\end{eqnarray}
where $X(t)=GC(t)\exp\{-\ii\Delta t \}$ and
$\displaystyle F^{\dagger}(t)\approx\sigma^+(0)+\sigma_z(0)C^{\dagger}(t){G\over\Delta}(1-\exp\{-\ii\Delta
t\})
$, 
if one takes into account the fact that the  modal dynamics is
very slow on the scale of the TLS dynamics and $\sigma_z(t)$ can
be considered as practically constant.  From Eq.
(\ref{jkmodel_solution_sigma_spon}), neglecting small and rapidly
oscillating terms,  after averaging over the atomic variables one
arrives to the following effective interaction Hamiltonian
\begin{eqnarray}
V_{JK}(t)\approx 2 {G^2\over\Delta}C^{\dagger}(t)C(t)+4
{G^4\over\Delta^3} C^{\dagger}(t)C(t)C^{\dagger}(t)C(t)+
{G\over\Delta}\left[C^{\dagger}(t)\Gamma(t)(1-\exp\{-i\Delta
t\})+\mbox{H. c.}\right]. \label{jkmodel_ham_V_app_spon}
\end{eqnarray}
Deriving the master equation in the standard manner, one obtains
an equation describing the correlated photon losses
\begin{eqnarray}
{\dd\over \dd t}\rho(t)\approx-\ii[\delta w C^{\dagger}C+\chi
C^{\dagger}CC^{\dagger}C ,\rho(t)]+{\bar\gamma}
\mathcal{L}(C)\rho(t), \label{master_jkmodel_spon}
\end{eqnarray}
where $\delta w=2{G^2/\Delta}$, $\chi=4{G^4/\Delta^3}$ and
${\bar\gamma}=2{\gamma G^2/\Delta^2}$.

Effectively, Eq.~(\ref{master_jkmodel_spon})
describe both coupling between modes and their interaction with
the same reservoir. Both these interaction might lead to the
entanglement between modes. As seen in Section V, even
in the absence of direct intermodal coupling (i.e. for $\delta
w=0$, $\chi=0$) an interaction of modes with the reservoir
entangles these modes.

The analysis made above can be readily generalized  to different schemes of cross-Kerr nonlinearity
generation through interaction of two modes with the same ensemble
of emitters \cite{{plenio08},{korolkova2007}}. In
Subsection~\ref{lambda-system}
we devise the procedure for the scheme of the giant cross-Kerr
nonlinearity generation suggested in Ref. \cite{plenio08}.


\subsection{An example: dispersive two-mode Jaynes-Cummings model
with  dephasing}

Here we illustrate an appearance of {\it correlated modal dephasing}
with the example of the Jaynes-Cummings model (Fig.~\ref{fig1})
considered in the previous Subsection. We model an influence of
the dephasing reservoir as a stochastic fluctuation of the TLS
transition frequency. In the rotating-wave approximation the problem
is described by the following Hamiltonian:
\begin{eqnarray}
H_{JK}=  w (a_1^{\dagger}a_1+a_2^{\dagger}a_2)+{1\over2}
[\omega_0+\zeta(t)]\sigma_z+
\left[\sigma^+(g_1a_1+g_2a_2)+(g_1a_1^{\dagger}+g_2a_2^{\dagger})\sigma^-\right],
\label{jkmodel_ham}
\end{eqnarray}
where  $\zeta(t)$ is random process describing a small rapid
stochastic modulation due to non-radiative interaction with
surroundings. For simplicity we take $\zeta(t)$ to be just a white noise
satisfying the following relations
\[ \langle\zeta(t)\rangle_{\rm s}=0, \quad
\langle\zeta(t)\zeta(\tau)\rangle_{\rm s}=d\,\delta(t-\tau),
\]
where  $\langle\ldots\rangle_{\rm s}$ denotes classical
averaging. We consider the case of the weak loss, $d \ll |g_{1,2}|$.

As before, we assume that the conditions for adiabatic
elimination of the emitter hold.  In the interaction picture with
respect to the part of the Hamiltonian (\ref{jkmodel_ham})
corresponding to the absence of the TLS-field interaction, we have
the following interaction Hamiltonian
\begin{eqnarray}
\label{jkmodel_ham_V} V_{JK}(t)=  G \left(\sigma^+C f(t)+\mbox{H.
c.}\right), \qquad f(t)=\exp\left\{\ii\Delta t+\ii\int\limits_0^t
\dd\tau \zeta(\tau)\right\}.
\end{eqnarray}
A formal solution for $\sigma^+(t)$ in this interaction picture
can be approximated as
\begin{eqnarray}
\sigma^+(t)\approx\sigma^+(0)+\ii G\int\limits_{0}^t\dd t_1
[2F^{\dagger}(t_1)F(t_1)-1]C^{\dagger}(t_1)f^*(t_1),
\label{jkmodel_solution_sigma_def}
\end{eqnarray}
where
\begin{eqnarray}
F^+(t)\approx\sigma^+(0)+\ii
G\sigma_z(0)C^{\dagger}(t)\int_0^t\dd\tau f^*(\tau).
\label{jkmodel_solution_sigma}
\end{eqnarray}

After averaging over TLS states, the following effective interaction
Hamiltonian can be obtained from Eq.~(\ref{jkmodel_solution_sigma}):
\begin{eqnarray}
V_{JK}(t)\approx  2 G^2C^{\dagger}(t)C(t)p^{(2)}(t)+4
{G^4\over\Delta^3} C^{\dagger}(t)C(t)C^{\dagger}(t)C(t),
\label{jkmodel_ham_V_app}
\end{eqnarray}
where
\begin{eqnarray}
\nonumber p^{(2)}(t)={\rm Re}\left[\ii f(t)\int_0^t\dd\tau f^*(\tau)\right].
\end{eqnarray}

Averaging over dephasing noise and neglecting small terms, we get the following master equation using
the standard technique implemented to derive Eq. (\ref{master_gener}):
\begin{eqnarray}
{\dd\over \dd t}\rho(t)\approx-\ii[\delta w C^{\dagger}C+\chi
C^{\dagger}CC^{\dagger}C,\rho(t)]+{\bar d}
\mathcal{L}(C^{\dagger}C)\rho(t), \label{master_jkmodel}
\end{eqnarray}
where
\begin{eqnarray}
{\bar d}=4G^4\left\langle \int\limits_0^{+\infty}\dd\tau
p(0)p(\tau)\right\rangle_{\rm s} \sim d{4G^4\over\Delta^4}.
\label{master_jkmodel_coefficients}
\end{eqnarray}
The calculations of the coefficients in (\ref{master_jkmodel_coefficients}) are carried out using the
following property \cite{{kilin1986},{gardiner}}:
\[\left\langle
\exp\left\{\ii\int\limits_0^t\dd\tau
\zeta(\tau)\right\}\right\rangle_{\rm s}=
\exp\left\{-\int\limits_0^t\dd\tau\int\limits_0^\tau \dd x
\left\langle\zeta(\tau)\zeta(x)\right\rangle_{\rm s}\right\}=\exp\left\{-dt\right\}
\]
and the fact that the detuning $\Delta$ is assumed to be
large, $\Delta \gg d$.

So, one can see that dephasing of the atom leads to appearance of
the correlated modal dephasing practically in the same manner as
atomic population losses lead to the correlated modal loss
considered in the previous Subsection. Also, modal dephasing
occurs notwithstanding the fact that the upper atomic level
remains practically unpopulated.


\begin{figure}[th]
\centerline{\psfig{file=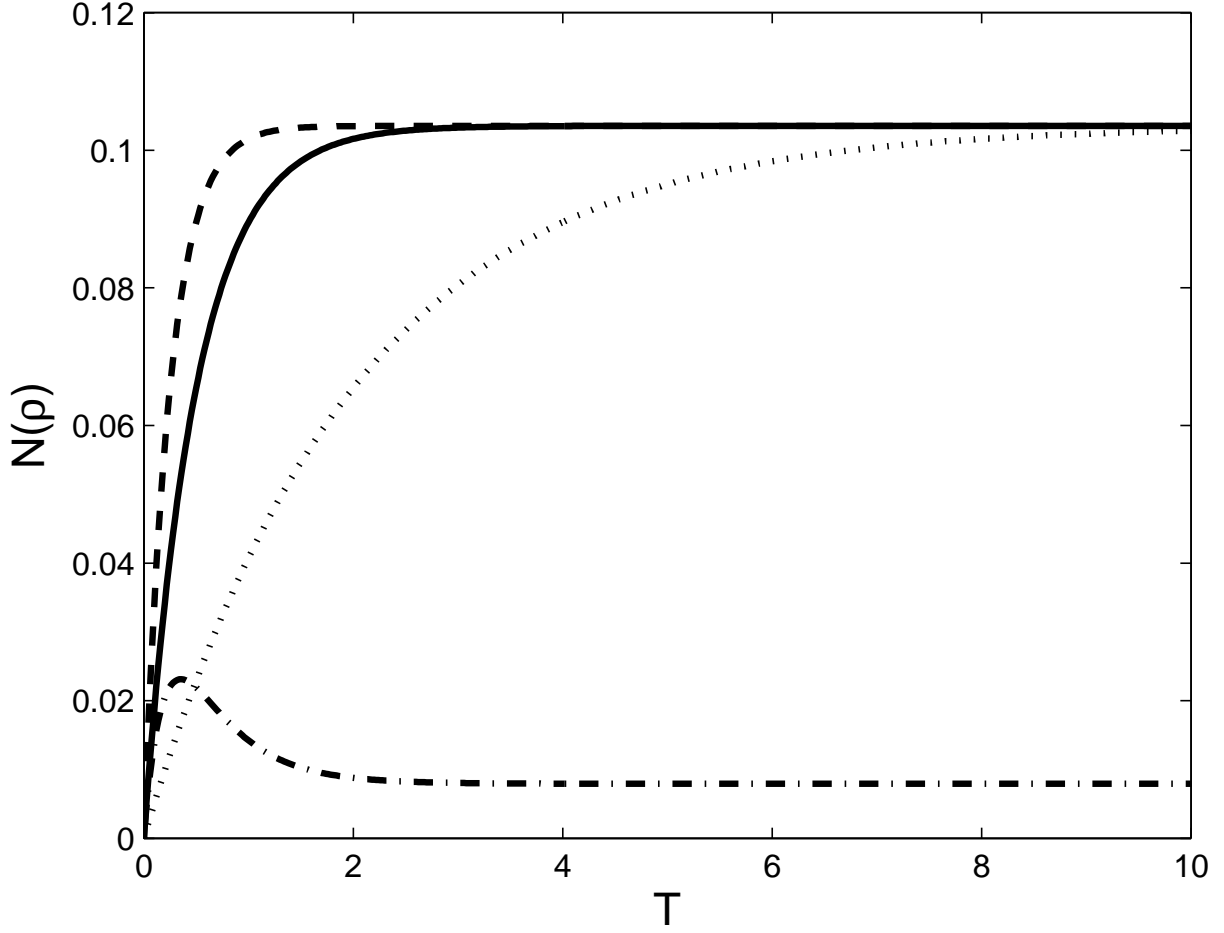,width=\linewidth}} \vspace*{8pt}
\caption{Examples of the negativity, $N(\rho)$,  dynamics given by
the solution (\ref{spon_system}) for the initially disentangled
state of modes $1$ and $2$ (namely, single photon in the mode $1$
and vacuum in the mode $2$).  The time, $T$, is given in units of
$g_2$; solid, dotted and dashed lines correspond to
${\bar\gamma}=g_2,0.25g_2,2g_2$ and $g_1=g_2$. Dash-dotted line
corresponds to ${\bar\gamma}=g_2$ and $g_1=2g_2$. Here
$\delta\omega+\chi=0$ for all graphs.} \label{fig5}
\end{figure}

\subsection{Beam-splitting by decoherence}

Finally, we demonstrate that the scheme described by the master
equation (\ref{master_jkmodel_spon}) can effectively produce
entanglement between the modes. In fact, this scheme can act as a kind of lossy
beam-splitter even in the absence of intermodal interaction in unitary part of
Eq.~(\ref{master_jkmodel_spon}). Indeed, let us consider a completely
uncorrelated single-photon initial state
$|\Psi(0)\rangle=|1\rangle_1|0\rangle_2$, and the initial density
matrix $|\rho(0)\rangle=|\Psi(0)\rangle|{\bar\Psi(0)}\rangle$ . In
the zero- and single-photon subspaces one can assume the following
orthonormal basis:
\[|\psi_{+}\rangle={1\over G}(g_1|1\rangle_1|0\rangle_2+
g_2|0\rangle_1|1\rangle_2), \quad |\psi_{-}\rangle={1\over
G}(g_2|1\rangle_1|0\rangle_2- g_1|0\rangle_1|1\rangle_2), \quad
|v\rangle=|0\rangle_1|0\rangle_2.
\]
One can easily see that the state $|\psi_{-}\rangle$ is not
affected by the losses described by Eq.
(\ref{master_jkmodel_spon}) because $C|\psi_{-}\rangle=0$. Also,
the following relations are satisfied:
\[C^{\dagger}C|\psi_{+}\rangle=|\psi_{+}\rangle, \quad
C|v\rangle=0.
\]
Thus the system of equations for the density matrix elements can be obtained from Eq.~(\ref{master_jkmodel_spon}):
\begin{eqnarray}
\nonumber \rho_{--}(t)=\rho_{--}(0), \quad {\dd\over
\dd t}\rho_{++}(t)=-2{\bar\gamma}\rho_{++}(t), \quad {\dd\over
\dd t}\rho_{+-}(t)=-[{\bar\gamma}+\ii (\delta w+\chi)]\rho_{+-}(t), \\
\rho_{-v}(t)=\rho_{-v}(0), \quad {\dd\over
\dd t}\rho_{+v}(t)=-[{\bar\gamma}+\ii (\delta w+\chi)]\rho_{+v}(t).
\label{spon_system}
\end{eqnarray}

The solution (\ref{spon_system}) describes an emergence of
entanglement form the initially uncorrelated state of both modes (
single photon in the mode $1$ and vacuum of the mode $2$). Figure
~\ref{fig5} depicts a measure of entanglement, a negativity as
given in Ref. \cite{negativity}
\[N(\rho)={1\over2}({\rm Tr}\sqrt{\sigma\sigma^{\dagger}}-1),
\]
where $\sigma$ is the density matrix $\rho$ partially transposed
with respect to the first mode. Non-zero value of the negativity
means that the state is the entangled one.  It can be seen that
the decay rate into the common reservoir does not affect the
finally reached entanglement. This rate affect only time during
which a stationary state is reached. It follows from this system
of equations that the initial uncorrelated state
$|1\rangle_1|0\rangle_2$ under the action of the correlated modal
loss asymptotically turns into
\begin{eqnarray}
|\rho(\infty)\rangle={g_2^2\over
G^2}|\psi_{-}\rangle|{\bar\psi_{-}}\rangle+\left(1-{g_2^2\over
G^2}\right)|v\rangle|{\bar v}\rangle.
\label{spon_entangled_state}
\end{eqnarray}
The state (\ref{spon_entangled_state}) is entangled for an
arbitrary $g^2_{1,2}>0$. However, the maximal degree of asymptotic
entanglement is reached when $g_1=g_2$, and with increasing of
difference between $g_1$ and $g_2$ the asymptotic entanglement
decreases (Figure~\ref{fig5}).

Note that the state (\ref{spon_entangled_state}) is  influenced
neither by the cross-Kerr interaction of the modes nor by the
linear excitation exchange, and one can set both $\delta
w=\chi=0$. The same type of state is produced by a correlated
dephasing described by Eq.~(\ref{master_jkmodel}). Effectively,
the correlation of reservoirs allows for existence of
decoherence-free subspaces to which the two-mode state eventually
evolves \cite{braun}.
Entangling through the common reservoir with appearance
of the long-living state similar to the one described by Eq.
(\ref{spon_entangled_state}) might occur for emitters and
collective reservoir modes near the band-edge in photonic crystals
\cite{mogilevtsev2005}.

\end{document}